\documentclass[10pt]{article}
\usepackage{geometry} 
\usepackage[usenames,dvipsnames]{color}
\usepackage{amssymb}
\usepackage{amsmath}
\usepackage{cite}

\usepackage{hyperref}

\hypersetup{
   colorlinks=true,         
   linkcolor=blue,          
   citecolor=blue,        
   urlcolor=Red            
}

\def\be{\begin{equation}}
\def\ee{\end{equation}}
\def\bea{\begin{eqnarray}}
\def\eea{\end{eqnarray}}

\def\1{\'{\i}}                           

\newcommand{\leftBR}[1]{\left\{\begin{aligned}#1\end{aligned}\right.}

\def\Q{Q}
\def\q{q}

\begin{document}

\title{\textbf{Extended noncommutative Minkowski spacetimes
 and hybrid gauge symmetries}}


\author{\textbf{Angel Ballesteros$^1$ and Flavio Mercati$^2$}
\vspace{12pt}
\\
$^1$Departamento de F\'isica, Universidad de Burgos,\\
E-09001 Burgos, Spain\\
$^2$ Dipartimento di Fisica, Universit\`a di Roma ``La Sapienza'',\\
P.le A. Moro 2, 00185 Roma, Italy;
\vspace{12pt}
\\
E-mail: $^1$\href{mailto:angelb@ubu.es}{angelb@ubu.es}; $^2$\href{mailto:flavio.mercati@gmail.com}{flavio.mercati@gmail.com}}

\maketitle

\begin{abstract}
We study the Lie bialgebra structures that can be built on the one-dimensional central extension of the Poincar\'e and (A)dS algebras in (1+1) dimensions. These central extensions admit more than one interpretation, but the simplest one is that they describe the symmetries of (the noncommutative deformation of) an Abelian gauge theory, $U(1)$ or $SO(2)$ on Minkowski or (A)dS spacetime. We show that this highlights the possibility that the algebra of functions on the gauge bundle becomes noncommutative. This is a new way in which the Coleman--Mandula theorem could be circumvented by noncommutative structures, and it is related to a mixing of spacetime and gauge symmetry generators when they act on tensor-product states. We obtain all Lie bialgebra structures on centrally-extended Poincar\'e and (A)dS which are coisotropic w.r.t. the Lorentz algebra, and therefore admit the construction of a noncommutative principal gauge bundle on a quantum homogeneous spacetime. It is shown that several different types of hybrid noncommutativity between the spacetime and gauge coordinates are allowed. In one of these cases, an alternative interpretation of the central extension leads to a new description of the well-known canonical noncommutative spacetime as the quantum homogeneous space of a quantum Poincar\'e algebra.
\end{abstract}


\section{Introduction}

Spacetime symmetries play an outstanding role in physics. Their effect on physical laws actually preceded the formulation of the notion of four-dimensional spacetime, and ultimately motivated it. In modern physics, although we know that General Relativity implies the breaking at large scales of Poincar\'e symmetry, the latter still plays a central role as the local symmetry group and it thus determines to a great extent the structure of the  possible Quantum Field Theory (QFT) candidates to describe the fundamental interactions and fields.
However, the application of standard QFT on Minkowski spacetime to quantum gravity encounters fundamental problems, and the assumption of local Poincar\'e (Lorentz) invariance might be at the core of the issue.  It appears highly significant, in this context, that the Coleman-Mandula no-go theorem~\cite{ColemanMandula,Fewster} implies that Poincar\'e symmetry cannot be generalized within the framework of QFT with Lie-group symmetries.


In the quest for the correct quantum theory of gravity we were given a few hints by nature, which may point us towards the appropriate mathematical structures that are needed to solve the problem. One of the most interesting ones comes from (2+1)-dimensional Quantum Gravity, a theory over which we have good control, see~\cite{AT,Witten1,Carlip2+1book}. Coupling this theory to matter fields and integrating away the gravitational degrees of freedom reveals that the effective background geometry that these fields see is not, as expected, Minkowski space (see~\cite{FreidelLivine} and references therein). Instead, one finds something that is best described as a `noncommutative spacetime', in the sense that the algebra of functions that intervene in the path integral is nonabelian~\cite{Matschull,FreidelLivine}. The symmetries of such noncommutative backgrounds are not described by Lie groups of isometries, but through a deformation thereof (the so-called quantum groups~\cite{drinfeld87, CP, MajidBook}), whose semiclassical counterparts (Poisson-Lie groups~\cite{DrinfeldPL,CP}) are also well-known to describe classical phase spaces of (2+1) gravity coupled to point particles (see~\cite{AMII,FR,MSquat,MS} and references therein).


Although in (3+1) dimensions it is not possible to reproduce the theoretical framework that allows to reach the above conclusions in (2+1) dimensions,  the notions of noncommutative spacetimes and their quantum-group symmetries are still entirely valid, and we can study the possible consistent choices for this noncommutativity. In particular, in the search for a `ground state' of (3+1) quantum gravity which plays the same role as the effective noncommutative spacetime of (2+1) quantum gravity, we are interested in \emph{maximally symmetric} noncommutative spacetimes, which admit a 10-dimensional algebra of symmetries  deforming the Poincar\'e algebra. A physically-relevant deformation should be controlled by a parameter with the dimensions of a length, which should be related with the Planck scale~\cite{MatteoFlavioBialgebras}. From this perspective, quantum deformations of the Poincar\'e Lie algebra and their Poisson-Lie counterparts have been thouroughly constructed and classified in the literature (for the (3+1) case and their associated noncommutative Minkowski spacetimes see~\cite{Lukierskia, Lukierskib, kMinkowski, BHOS4D, tmatrix, nullplane, Zakrzewski1997, Lukitwists} and references therein). 

In this paper we intend to investigate those  features of quantum Poincar\'e groups and algebras  that could allow to overcome the limitations imposed by the no-go theorems on Lie-group symmetries of QFT. This is the fact, stated by the Coleman--Mandula theorem~\cite{ColemanMandula} (hereafter CMT), that in ordinary QFT it is impossible to  ``combine space-time and internal symmetries in any but a trivial way''. With ``trivial way to combine symmetries'' Coleman and Mandula mean a direct product of groups, and thus this theorem precludes the existence of the so-called ``hybrid symmetries"~\cite{Mandula} in which the space-time and the gauge symmetries are intertwined. Supersymmetry offers a way to circumvent the no-go theorem by making use of graded symmetry algebras which introduce fermionic generators~\cite{HLS}. Quantum group and Yangian symmetries arising from the Hopf algebras of nonlocal conserved quantities of certain integrable (1+1) field theories were soon proven to avoid the Coleman-Mandula constraint~\cite{BernardLeclair}, although this result cannot be generalized to (3+1) dimensions. On the other hand, the construction of QFT theories on twisted canonical noncommutative spacetimes constitutes another remarkable attempt in the same direction (see~\cite{Vassilevich, Aschieri, CT, CTZ} and references therein). 
Also, space-like extra dimensions have been recently proposed as a novel mechanism to connect internal and spacetime symmetries~\cite{LO}. 

In this work we adopt a purely kinematical perspective, and we focus on the quantum group deformations of the Poincar\'e symmetries in order to show that they could allow to circumvent the CMT from a different viewpoint.
In fact one of the fundamental assumptions of the CMT is a Leibniz action of the Poincar\'e algebra generators $T$ on  tensor product states: 
$$
T \triangleright |\psi \rangle \otimes  |\phi \rangle  = ( T \triangleright  |\psi \rangle) \otimes  |\phi \rangle  + |\psi \rangle \otimes  ( T \triangleright |\phi \rangle).
$$
One of the features of quantum groups is that, because of the noncommutativity of the functions over the group, the quantum algebra generators act in a nonlinear way on product states 
$$
T \triangleright |\psi \rangle \otimes  |\phi \rangle  = \sum_i ( T^{(1)}_i \triangleright  |\psi \rangle) \otimes  ( T^{(2)}_i \triangleright|\phi \rangle),
$$
as encoded by the `coproduct' map $\Delta$ (see below) which defines the way in which representations on tensor product states are defined.
In the simplest case of an Abelian internal symmetry, the spacetime symmetry group is enlarged through a \emph{central extension} $Q$. The essential result to be emphasized is that centrally extending a quantum-group deformation of a spacetime symmetry group gives, in general, a different result than deforming the centrally-extended group in its entirety. In other words, the operations of quantum deformation and central extension do not commute. The same holds, more in general, for taking the direct product of two groups: their quantum group deformation does not have to be the direct product of two quantum groups. In this case too it is the `coproduct' structure that allows for more freedom on how algebra generators act on tensor products. As we will see in the sequel, [see Eq.~\eqref{Pre-cocommutator_TriviallyExtendedPoincare}] under quantum deformations one can imagine a gauge generator $\Q$  mixing with Poincar\'e generators $T_i$ when acting on tensor products, \emph{i.e.}
\begin{equation*}
\begin{aligned}
&\Q \triangleright |\psi \rangle \otimes  |\phi \rangle  = \Q \triangleright  |\psi \rangle \otimes  |\phi \rangle  + |\psi \rangle \otimes   \Q \triangleright |\phi \rangle
\\
&+ c^{(1)}_{ij}  T_i \triangleright  |\psi \rangle \otimes    T_j \triangleright |\phi \rangle 
+ c^{(2)}_i Q \triangleright  |\psi \rangle \otimes    T_i \triangleright |\phi \rangle 
+ c^{(3)}_i T_i \triangleright  |\psi \rangle \otimes    Q \triangleright |\phi \rangle 
 + c^{(4)} Q \triangleright  |\psi \rangle \otimes    Q \triangleright |\phi \rangle + \dots \,,
 \end{aligned}
\end{equation*} 
 or the opposite: 
\begin{equation*}
\begin{aligned}
&T_i \triangleright |\psi \rangle \otimes  |\phi \rangle  =  T_i \triangleright  |\psi \rangle) \otimes  |\phi \rangle  + |\psi \rangle \otimes  ( T_i \triangleright |\phi \rangle) 
\\
&+ d^{(1)}_{ijk}  T_j \triangleright  |\psi \rangle \otimes    T_k \triangleright |\phi \rangle 
+ d^{(2)}_{ij} Q \triangleright  |\psi \rangle \otimes    T_j \triangleright |\phi \rangle 
+ d^{(3)}_{ij} T_j  \triangleright  |\psi \rangle \otimes    Q \triangleright |\phi \rangle 
 + d^{(4)} Q \triangleright  |\psi \rangle \otimes    Q \triangleright |\phi \rangle + \dots
 \,,
 \end{aligned}
\end{equation*} 
where $c^{(a)}_{i\dots}$ and  $d^{(a)}_{i\dots}$ are numerical coefficients, and an expansion in powers of the generators $Q$ and $T_i$ is understood.
In fact, this paper presents all possible hybrid actions of this type that are allowed by quantum symmetries in (1+1) dimensions.

This is an intriguing possibility, suggesting that in a noncommutative spacetime gauge transformations of composite fields might involve a small component of \emph{e.g.} translation, or Lorentz transformation. If the central generator $\Q$ is interpreted as the generator of Abelian $SO(2)$ gauge transformations, then the dual coordinate $\q$ will represent a coordinate on a neighbourhood of the identity of the  $SO(2)$ group manifold, which is $S^1$. Therefore, our approach intends to explore the possibility of making the manifold $\mathbb{M}\times S^1$ (the Cartesian product of Minkowski space $\mathbb{M}$ with $S^1$, usually interpreted as a trivial fibre bundle) noncommutative, and to get an explicit description of all possible roles played by the gauge coordinate in its interplay with the spacetime ones.
In this sense, the possibility explored in the present paper can offer a non-trivial way to unify gauge (`internal') and spacetime (`external') symmetries: the basic geometric object of field theory, the (trivial) Cartesian product of Minkowski space with the internal space, could be replaced by a unified object in which it is not always possibe to distinguish the spatial `external' directions from the gauge `internal' ones.

The natural starting point of such an investigation is the (1+1) dimensional Poincar\'e group with a one-dimensional central extension, which in the commutative case represents the symmetries of Abelian gauge theory, but we stress that the approach here presented is fully applicable in any dimension and also for higher dimensional (and non-abelian) gauge groups. A systematic characterization of quantum group deformations of the centrally extended Poincar\'e algebras is -- to the best of our knowledge -- so far an unexplored issue. Some past works considered central extensions of the Poincar\'e algebra in the context of quantum groups, \emph{e.g.}~\cite{BHOS2D,azcarrag}, but only some isolated cases were considered and the possible interpretation of central extensions as gauge symmetries was not developed. Moreover, in the classification here presented we will take into account that Minkowski spacetime is obtained as the homogeneous space associated to the quotient of the Poincar\'e group by the Lorentz subgroup (which is the `isotropy subgroup', \emph{i.e.} the subgroup that leaves a point of the homogeneous spacetime invariant). Since we are interested in the noncommutative spacetimes whose symmetries are described by our quantum-deformed group, it turns out that requiring that the isotropy subgroup closes a sub-Hopf algebra under deformation is too strong a requirement (see, for instance,~\cite{Koor} and references therein). In fact, the appropriate notion is that of \emph{coisotropy} of the cocommutator of the Lie bialgebra on the isotropy subalgebra~\cite{Zakrzewski,BMN} (see below). Essentially, this condition amounts to asking that the translation coordinates on the group close an algebra, and this algebra can be taken as the definition of a noncommutative algebra of coordinates on the quantum homogeneous spacetime. The coisotropy condition will be used througout this paper as the essential constraint that limits the number of possible Lie bialgebra structures (and consequently, of quantum deformations) for the centrally extended Poincar\'e algebra. Moreover, we will show that the very same classification scheme can be applied to the centrally extended (anti) de Sitter algebra and its associated extended noncommutative spacetimes.

The structure and main results of the paper are the following. In Sec.~\ref{SecPreliminaries} we present the basics of quantum groups, Lie bialgebras and their associated first-order noncommutative spacetimes, which is illustrated with the example of the (non-extended) Poincar\'e algebra in (1+1) dimensions. In this case the so-called `$\kappa$-Minkowski' noncommutative spacetime is selected by the coisotropy condition as the unique possible quantum homogeneous space. The detailed analysis of the centrally extended Poincar\'e Lie bialgebras constitutes the core of the paper, and is given in  Sec.~\ref{Sec1+1trivialExtensionPoincare}. Here all possible types of (1+1) ``hybrid gauge symmetries'' and, therefore, of their associated noncommutative algebras generated by the spacetime, $x$, and gauge, $q$ coordinates are obtained as the quantum homogeneous spaces arising from coisotropic quantum deformations of the centrally extended Poincar\'e Lie algebra. One of the cases so obtained is just the so-called canonical or $\theta$-noncommutative spacetime (see~\cite{DFR, szabo, Aschieri2017} and references therein), that can be thus interpreted as being the quantum homogeneous space of a certain quantum Poincar\'e group. In Sec.~\ref{Sec(A)dS} we consider the Lie bialgebra structures for the central extension of the (1+1) dimensional (Anti-) de Sitter  algebra. We find that the (A)dS case also allows the introduction of hybrid gauge symmetries for a nonvanishing cosmological constant, but in general they turn out to be more restrictive than the Poincar\'e ones (for instance, the $\theta$-noncommutativity cannot be recovered). In Sec.~\ref{SecCanonical} 
an application of the approach here presented in higher dimensions is given by studying the possibility of extending the interpretation of $\theta$-noncommutativity as quantum homogeneous space to dimensions higher than (1+1) is analysed, but the result is negative. Finally, in Sec.~\ref{SecConclusions} we summarize and discuss the results presented in the paper, and we comment on several future research problems.


\section{From Lie bialgebras to noncommutative spacetimes}\label{SecPreliminaries}

In this section we recall how first-order noncommutative spacetimes are obtained from their Lie bialgebra of symmetries. The constraints imposed onto a Lie bialgebra by the request of underlying a quantum homogeneous space are also summarized. A detailed description of quantum groups, Lie bialgebras and Poisson/quantum homogeneous spaces can be found in~\cite{drinfeld87, CP, MajidBook, DrinfeldPL, Koor, Zakrzewski, BMN} and references therein.


\subsection{Lie bialgebras}

A Lie bialgebra $(g,\delta)$ is Lie algebra $g$ with structure tensor $c^k_{ij}$ 
\be
[X_i,X_j]=c^k_{ij}X_k ,
\label{liealg}
\ee
which is also endowed with a skew-symmetric ``cocommutator'' map
$$
\delta:{g}\to {g}\wedge {g}
$$
fulfilling the two following conditions:
\begin{itemize}
\item i) $\delta$ is a 1-cocycle, {\em  i.e.},
$$
\delta([X,Y])=[\delta(X),\,  Y\otimes 1+ 1\otimes Y] + 
[ X\otimes 1+1\otimes X,\, \delta(Y)] ,\qquad \forall \,X,Y\in
{g}.
\label{1cocycle}
$$
\item ii) The dual map $\delta^\ast:{g}^\ast\otimes {g}^\ast \to
{g}^\ast$ is a Lie bracket on ${g}^\ast$.
\end{itemize}
As a consequence, any cocommutator $\delta$ will be of the form
\be
\delta(X_i)=f^{jk}_i\,X_j\wedge X_k \, ,
\label{precoco}
\ee
where $f^{jk}_i$ is the structure tensor of the dual Lie algebra $g^\ast$ defined by
\be
[ \xi^j, \xi^k]=f^{jk}_i\, \xi^i \, ,
\label{dualL}
\ee
where $\xi^l$ are the dual generators, defined by the pairing $\langle   \xi^j,X_k \rangle=\delta^j{}_k$. If, as usual in Lie group theory, we interpret this pairing as the duality between a Lie algebra generator and its corresponding local coordinate around the identity of the Lie group $G$, then relations~\eqref{dualL} invite the interpretation of $\xi_i$ as the noncommutative group coordinates. In particular, under certain conditions, a subset of these coordinates can be identified with the ``local'' coordinates of a given spacetime obtained as $G/H$ with $H$ being some isotropy subgroup.

The problem of obtaining all possible Lie bialgebra structures $(g,\delta)$ on a given  Lie algebra $g$ can be addressed by
 first solving the 1-cocycle condition, namely,
\be
f^{ab}_k c^k_{ij} = f^{ak}_i c^b_{kj}+f^{kb}_i c^a_{kj}
+f^{ak}_j c^b_{ik} +f^{kb}_j c^a_{ik},
\label{compatfc}
\ee
where the $c^k_{ij}$ tensor is known and linear equations~\eqref{compatfc} have to be solved for the $f^{ab}_k$ tensor. Second, the Jacobi identities of $g^\ast$ impose quadratic equations onto the components of $f^{ab}_k$ that have not yet been fixed by~\eqref{compatfc}. Therefore, a given finite dimensional Lie algebra $g$ could admit a large number of Lie bialgebra structures $\delta$, whose equivalence classes can be computed by making use of the automorphisms of $g$. Also, as it could be expected, for some Lie bialgebras the 1-cocycle $\delta$ is a coboundary
\be
\delta(X)=[ X \otimes 1+1\otimes X ,\,  r],\qquad 
\forall\,X\in {g} ,
\label{cocom}
\ee
where $r$ is a skew-symmetric element of
${g}\otimes {g}$ given by 
$
r=r^{ab}\,X_a \wedge X_b\, ,
$
which in order to define an admissible $\delta$  has
to be a solution of the modified classical Yang--Baxter equation (mCYBE)~\cite{CP, MajidBook}.
For semisimple Lie algebras all Lie bialgebra structures are coboundaries, and that is also the case for the Poincar\'e algebra in (2+1) and (3+1) dimensions, despite these two Lie algebras are not semisimple~\cite{Zakrzewski1997}. In the case of Lie algebras  with central generators, non-coboundary Lie bialgebra structures arise naturally, and only a fraction of the Lie bialgebras that can be built on them admit a classical $r$-matrix (see for instance~\cite{EnricoGalilei,Opanowicz21} where the case of the centrally extended Galilei algebra is studied, or~\cite{heis} where the Lie bialgebra structures on the Heisenberg-Weyl algebra were classified and the corresponding quantum Hopf algebras were explicitly constructed). Since in our approach to first-order noncommutative spacetimes we will not make use of $r$-matrices, we will omit them in the sequel, although they can be explicitly found by solving the linear equations~\eqref{cocom} for the those of the Lie bialgebras here presented which are coboundaries.


\subsection{Quantum groups/algebras, Poisson-Lie groups and Lie bialgebras}

A quantum group is an algebraic generalization of the notion of Lie group where the algebra of functions on the group manifold is replaced by a noncommutative algebra for which the group multiplication is an algebra homomorphism. To describe such a structure, we need the language of Hopf algebras: a unital algebra $H$ on which, in addition to a product map $\cdot : H \otimes H \to H$, a `coproduct' map $\Delta : H \to H \otimes H$, together with an `antipode' $S: H \to H$ and a `counit' $\varepsilon : H \to \mathbb{C}$ are defined. These three additional maps have to be (anti) homomorphisms w.r.t.~the algebra product, and they are used to encode the group structures. For example, in the case of the general linear group $GL(n)$, the three maps read
\begin{equation}
\Delta (M^i{}_j) = M^i{}_k \otimes M^k{}_j \,, \qquad S(M^i{}_j) = (M^{-1})^i{}_j \,, \qquad \varepsilon(M^i{}_j) = \delta^i{}_j \,.
\end{equation}
The three maps have to be consistent with the group axioms: associativity, the existence of the identity and the existence of the inverse. These translate into a \emph{coassociativity} property of $\Delta$, and an identity involving the three maps and the multiplication: $\cdot \circ ( S \otimes  \text{id} ) \circ \Delta = \cdot \circ ( \text{id} \otimes S ) \circ \Delta = 1 \,\epsilon$. These rules, together with the homomorphism property of the three maps, make a consistent algebraic structure. We can describe in this way a standard Lie group in terms of the algebra of functions on the group manifold, if the product $\cdot$ is assumed commutative, but we can also generalize to the case in which $\cdot$ is noncommutative: in that case the Hopf algebra still makes sense, but it does not describe an ordinary Lie group. We are in this case dealing with a genuine quantum group.

To generalize the (extremely useful) notion of Lie algebra, it is sufficient to take the Hopf algebra dual $H^*$ of a quantum group, which is by definition a Hopf algebra. Here the algebra elements are not functions on a group manifold: they are elements of the \emph{universal enveloping algebra} $U(g)$ of the Lie algebra $g$ associated to the group. The dual product map in this case encodes the commutation rules of the Lie algebra. In the example of $GL(n)$,the Lie algebra generators $t_i$ are again $n\times n$ matrices, but the Lie algebra product is defined by the matrix commutator. The coproduct, antipode and counit in this case define the Lie algebra generators as differential operators on the Lie group:
\begin{equation}
\Delta (t_i)= t_i \otimes 1 + 1 \otimes t_i \,, \qquad S(t_i) = - t_i \,, \qquad \varepsilon(t_i) = 0 \,.
\end{equation}

All of these maps will be deformed in a nonlinear way in the noncommutative case. This is why the notion of quantum universal enveloping algebra, which admits nonlinear commutation rules between the generators, arises. In this context, the first order deformation of a quantum algebra provides the relevant information concerning the type of quantum deformation we are constructing. Consider a quantum algebra generated by $t_i$, and an expansion of the commutators and the coproducts in powers of the generators:
\begin{equation}
[ t_ i , t_j ] = c_{ij}{}^k t _k + c'_{ij}{}^{kl} t_k \, t_l + \mathcal{O}(t^3) \,, ~~ 
\Delta (t_i) = t_i \otimes 1 + 1 \otimes t_i + f_i{}^{jk} \, t_j \otimes t_k + \mathcal{O}(t^2) \,.
\end{equation}
The constants $c_{ij}{}^k$ have to be structure constants of a Lie algebra: in fact the product of the universal enveloping algebra is associative, and this implies that $c_{ij}{}^k$ satisfies the Jacobi identities. Similarly, since the coproduct $\Delta$ has to be coassociative and it is a homomorphism with respect to the product, the dual map defined by the antisymmetric (in $j$, $k$) part of  $f_i{}^{jk}$ satisfies the Jacobi identities and the 1-cocycle condition~\eqref{compatfc}. Therefore \emph{the first-order term of the power expansion of a quantum algebra is a Lie bialgebra}. The opposite can usually be proved too: starting with a Lie bialgebra and requiring the homomorphism and cocycle conditions order by order, we can define an associated quantum algebra as a power series in the generators. This highlights the importance of Lie bialgebras for quantum algebras. 

The nature of the difficulties that one encounters in going from the Lie bialgebra to the quantum algebra are best illustrated by considering the relation with quantum groups. A quantum group can be understood as a \emph{quantization} of a Poisson-Lie group. This is a Lie group on which a Poisson structure (\emph{i.e.}~a Poisson bracket) is defined, as an operation on the algebra of functions on the group, which is compatible with the group operations (\emph{i.e.}~they must be \emph{Poisson maps}). The Poisson bracket on the Poisson-Lie group is the classical precursor of the noncommutative product between functions on the associated quantum group. The procedure of `quantization' is not unique because of ordering ambiguities could appear due to the fact that Poisson-Lie structures are, in general, nonlinear in terms of the local coordinates on the group. Lie bialgebras enter this picture because they are the Lie algebras of Poisson-Lie groups and are in one-to-one correspondence with them: the first-order expansion of the Poisson bracket gives the (structure constants of a) cocommutator of the Lie bialgebra, and the fact that the product is a Poisson map implies the 1-cocycle condition. Therefore, given a Lie algebra $g$, the knowledge of its Lie bialgebra structures $(g,\delta)$ provides the ``directions'' along which we can obtain its quantum group and quantum algebra deformations.

\subsection{Poisson/quantum homogeneous spaces}

%
%
%
%
%

In the commutative case, a homogeneous space $M$ is obtained as the quotient
\be
M=G/H \,,
\ee
where $G$ is the group of isometries of the spacetime [\emph{e.g.} $ISO(3,1)$ (Poincar\'e), $SO(4,1)$ (AdS) or $SO(3,2)$ (AdS)] and $H$ is the invariance subgroup of a point of the space (the Lorentz subgroup $SO(3,1)$). If we are dealing with a quantum group, we need to generalize this notion so that the isometry group leaves the noncommutative structures of the homogeneous space invariant. In~\cite{BMN} such a generalization was discussed at the level of Poisson-Lie groups. A `Poisson homogenous space' $(M,\pi)$ is a quotient $M=G/H$, together with a Poisson bracket $\pi$ which is invariant under the action of $G$ endowed with the Poisson-Lie structure $\Pi$. This implies that for a given $\Pi$ the corresponding $\pi$ has to be found, and this turns out to be guaranteed if the Lie bialgebra $(g,\delta)$ associated to $\Pi$ fulfills the following `coisotropy' condition:
\be\label{coisotropy}
\delta (Y) \in Y \wedge X \,, \qquad Y \in h \,, ~~ X \in g \,,
\ee
where $h$ is the Lie algebra of the isotropy subgroup $H$. 
Since the duals of the generators of the Lie bialgebra $\langle   \xi^j,X_k \rangle=\delta^j{}_k$ can be interpreted as `infinitesimal' coordinates on a neighbourhood of the origin of the quantum group, then their Lie brackets~\eqref{dualL} provide a first-order expansion of the commutation relations of the nonabelian algebra of functions on the group. Then, if we call $y$ the generators dual to $Y \in h$,  Eq.~\eqref{coisotropy} states that the $y$ generators can only appear on the right-hand side of the Lie brackets between $y$ and any other generators. Therefore, if we identify a linear complement of $h^*$, its generators $x$ will close a Lie subalgebra. Moreover, we can identify this subalgebra as the noncommutative algebra of coordinates on the quantum homogeneous space. This is the ultimate meaning of coisotropy: asking that the coordinates on the noncommutative space close algebraically.

Since we will be interested in constructing noncommutative spacetimes associated with quantum group deformations, the rest of the paper we will be dealing with coisotropic Lie bialgebras, which will define for our purposes the relevant subset of physically relevant quantum groups. Our Lie bialgebras will be based on a central extension of the Poincar\'e and (A)dS algebras in (1+1) dimensions, which means $g = iso(1,1) \oplus  so(2)$ in the Poincar\'e case and $g = so(2,1) \oplus  so(2)$ in the case of (A)dS. As isotropy subalgebras we can either choose the Lorentz subalgebra $so(1,1)$ or the direct sum of the Lorentz and $so(2)$ subalgebras, $so(1,1)\oplus so(2)$. In the interpretation of $G$ representing the symmetries of a gauge bundle on a homogeneous spacetime, the first case corresponds to quotenting down to the bundle, while the second case corresponds to quotienting out the gauge coordinate as well and projecting down to spacetime. We will see in what follows, however, that this interpretation is not tenable in all cases.


\subsection{(1+1) Poincar\'e Lie bialgebras}

As a warmup, we will compute the most generic Lie bialgebra structure on the Poincar\'e algebra in (1+1) spacetime dimensions, $iso(1,1)$, and we will impose onto it the coisotropy condition for the Lorentz subalgebra.
The $iso(1,1)$ Lie algebra and its quadratic Casimir are
\be
[K,P_0]=P_1 \,, \qquad [K,P_1]=P_0 \,, \qquad [P_0,P_1]=0  \,,
\label{Poincare1+1_alg}
\ee
\be
{\cal C}=P_0^2 - P_1^2 \,,
\label{Poincare1+1_cas}
\ee
where $P_0$ and $P_1$ are the generator of time- and space-like translations and $K$ is the generator of boosts. To find the most general Lie bialgebra structure $(iso(1,1),\delta)$ we write a generic (pre-)cocommutator~\eqref{precoco}, and impose the 1-cocycle~\eqref{compatfc} and Jacobi conditions on the structure tensor $f_i^{jk}$. These equations admit two independent solutions:
\be
\left\{
\begin{aligned}
&\delta(K)=  K \wedge \left( a^0 \, P_0  + a^1 \,  P_1 \right)   \,, \cr
&\delta(P_0)= a^1 \,P_0\wedge P_1   + b \,K \wedge P_0 \,, \cr
&\delta(P_1)=  a^0 \,P_1\wedge P_0  + b \,K \wedge P_1 \,,
\end{aligned} \right.
\qquad
\left\{
\begin{aligned}
&\delta(K)=  K \wedge \left( a^0 \, P_0 + a^1 \,  P_1 \right) +  c \,P_1\wedge P_0 \,, \cr
&\delta(P_0)=  a^1 \,P_0 \wedge P_1  \,,\cr
&\delta(P_1)=  a^0 \,P_1\wedge P_0 \,. \cr
\end{aligned}\right.
\ee
depending each of them on three real parameters: $(a^0,a^1,b)$ and $(a^0,a^1,c)$.  

Since we will be interested in the (1+1) noncommutative Minkowski spacetimes arising from $H$ being the Lorentz subgroup, the coisotropy condition w.r.t. the Lorentz subalgebra reads:
\be
\delta(K)\subset K\wedge X,  \qquad X\in {g} \,.
\label{so11coisotropy}
\ee
It is immediate to check that this condition generates the constraint $c=0$ on the second solution, which now is converted into the $b=0$ subcase of the first family.
We thus conclude that the most general bialgebra deformation of $iso(1,1)$ which is coisotropic w.r.t. the Lorentz subalgebra is
\bea\label{cocosinm}
\leftBR{
& \delta(K)=  K \wedge \left( a^0 \, P_0 + a^1 \,  P_1 \right)   \,, 
\\
& \delta(P_0)= a^1 \,P_0 \wedge P_1   + b \,K \wedge P_0 \,,
\\
&\delta(P_1)=  a^0 \,P_1\wedge P_0  + b \,K \wedge P_1 \,.
}
\eea
Introducing a dual basis for $g^\ast$ through the pairing
\be \label{1+1DualBasis}
\begin{aligned}
&\langle x^0 , P_0 \rangle = 1  \,, \qquad  &\langle x^1 , P_0 \rangle = 0 \,, \qquad &\langle \chi , P_0 \rangle  =1 \,, \\
&\langle x^0 , P_1 \rangle =  0 \,, \qquad &\langle x^1 , P_1 \rangle = 1 \,, \qquad  &\langle \chi , P_1 \rangle  =0 \,,\\
&\langle x^0 , K \rangle = 0 \,, \qquad &\langle x^1 , K \rangle = 0 \,, \qquad  &\langle \chi , K \rangle  =1 \,,
\end{aligned}
\ee
the Lie algebra dual to the coisotropic solution~\eqref{cocosinm} is obtained by dualizing $\delta$, namely:
\be
[x^0 , x^1 ] = a^1 \, x^0 - a^0  \, x^1 \,, \qquad [\chi , x^0 ] =  b  \,  x^0 + a^0 \,  \chi \,, \qquad  [\chi , x^1 ] =  b  \, x^1 + a^1 \,  \chi \,.
\ee
These expressions provide the most generic first-order noncommutative Minkowski spacetime that can be obtained from quantum deformations of the (1+1) Poincar\'e algebra and satisfy the coisotropy condition. The algebra of functions on this noncommutative spacetime is generated by the subalgebra $(x^0,x^1)$:
\be\label{genkm11}
[x^0 , x^1 ] = a^1 \, x^0 + a^0  \, x^1  \, ,
\ee
and $x^0$, $x^1$ admit the interpretation of noncommutative coordinate functions.
The parameters $a^0$ and $a^1$ represent the components of a vector which generalizes the well-known $\kappa$-Minkowski noncommutative spacetime~\cite{kMinkowski}, which is straightforwardly recovered in the  time-like case $a^0=1/\kappa$, $a^1 =0$ (the light-cone quantum deformation~\cite{tmatrix} is also obtained when $a^0=\pm\,a^1$). Moreover, it is easy to check that the algebra~\eqref{genkm11} is isomorphic to the $a^1=0$ case provided that $a^0 \neq 0$. $\kappa$-Minkowski turns out to be the only noncommutative Minkowski spacetime that can be obtained as the homogeneous space of a quantum group. As we will see in the sequel, this will drastically change when quantum deformations of a central extension of the Poincar\'e algebra are considered.

\section{Extended (1+1) noncommutative Minkowski spacetimes}\label{Sec1+1trivialExtensionPoincare}

The aim of this section is to perform the same Lie bialgebra analysis for the case of the one-dimensional  central extension of the (1+1)  Poincar\'e Lie algebra, which is defined by the Lie brackets
and Casimir operators 
\be
[K,P_0]=P_1 \,, \qquad [K,P_1]=P_0 \,,  \qquad [P_0,P_1]=0 \,, \qquad [\Q,\cdot\,]=0 \,,
\label{bcext}
\ee
\be
{\cal C}_1= \Q, \qquad {\cal C}_2=P_0^2 - P_1^2 \,.
\label{bdext}
\ee
The generator $\Q$ defines a one-dimensional abelian subalgebra, which is often called a pseudo-extension or `trivial' central extension, due to the fact that the extended Lie algebra can be written as a direct sum of the initial Lie algebra plus the extra generator (see~\cite{azcarrag} and references therein). We also recall that the algebra~\eqref{bcext} plays a relevant role in (1+1) gravity and is sometimes called the Nappi-Witten algebra (see~\cite{NappiWitten,CGannals}). Note that the dual Lie algebra $g^\ast$ contains a fourth generator $\q$ which, in the interpretation of~\eqref{bcext} representing the symmetries of a $U(1)$ gauge bundle on Minkowski spacetime, would stand for the gauge coordinate.


\subsection{The general solution}

The most general pre-cocommutator on~\eqref{bcext} depends on $24$ real parameters
\be
\label{Generalcocommutator_TriviallyExtendedPoincare}
\leftBR{
&\delta(K)= k_1 ~K\wedge P_1+  k_2 ~K\wedge P_0 + k_3 ~K\wedge \Q + k_4 ~P_1\wedge P_0 + k_5 ~P_1\wedge \Q +  k_6 ~P_0\wedge \Q  \,, \cr
&\delta(P_0)= h_1 ~K\wedge P_1+  h_2 ~K\wedge P_0 + h_3 ~K\wedge \Q + h_4 ~P_1\wedge P_0 + h_5 ~P_1\wedge \Q +  h_6 ~P_0\wedge \Q  \,, \cr
&\delta(P_1)= p_1 ~K\wedge P_1+  p_2 ~K\wedge P_0 + p_3 ~K\wedge \Q + p_4 ~P_1\wedge P_0 + p_5 ~P_1\wedge \Q +  p_6 ~P_0\wedge \Q \,, \cr
&\delta(\Q)= m_1 ~K\wedge P_1+ m_2 ~K\wedge P_0 + m_3 ~K\wedge \Q + m_4 ~P_1\wedge P_0 + m_5 ~P_1\wedge \Q +  m_6 ~P_0\wedge \Q  \,.
}
\ee
By imposing the cocycle condition we go down to 14:
\be\label{Pre-cocommutator_TriviallyExtendedPoincare}
\leftBR{
&\delta(K)= k_1 ~K\wedge P_1+  k_2 ~K\wedge P_0  + k_4 ~P_1\wedge P_0 + k_5 ~P_1\wedge \Q +  k_6 ~P_0\wedge \Q  \,, \cr
&\delta(P_0)=  h_2 ~K\wedge P_0  -k_1 ~P_1\wedge P_0 + h_5 ~P_1\wedge \Q +  h_6 ~P_0\wedge \Q  \,, \cr
&\delta(P_1)= h_2 ~K\wedge P_1  + k_2 ~P_1\wedge P_0 + h_6 ~P_1\wedge \Q +  h_5 ~P_0\wedge \Q \,, \cr
&\delta(\Q)=  m_4 ~P_1\wedge P_0  \,, 
}
\ee
and by enlarging the  dual basis (\ref{1+1DualBasis}) with the dual $\q$ of the central generator $Q$
\bea
\langle \q , P_0 \rangle = \langle \q , P_1 \rangle = \langle \q , K \rangle = 0 \,, ~~  \langle \q , \Q \rangle =1 \,,
\eea
the pre-cocommutator gives rise to the following Lie bracket:
\be
\begin{aligned}
& [x^0,x^1]= k_1 \,x^0 -  k_2 \,x^1- k_4  \chi - m_4 \,\q  \,,    \cr
& [x^0,\chi]=- h_2 \,x^0 - k_2 \,\chi  \,, \cr
& [x^0,\q]=  h_5  \, x^1+  h_6 \, x^0+  k_6 \,\chi  \,,
\end{aligned}
\begin{aligned}
& [x^1,\chi]= - h_2 \,x^1 -  k_1 \,\chi   \,,\cr
&  [x^1,\q]=  h_5 \, x^0 +  h_6 \,x^1+ k_5 \,\chi  \,, \cr
&   [\chi,\q]= 0  \,.
\end{aligned}
\label{dual11}
\ee
Finally, by imposing Jacobi identities onto~\eqref{dual11} we get the set of nonlinear equations:
\be
\begin{aligned}
& h_2 \, k_4 =0  \,, \cr 
& h_2 \,  m_4 =0   \,,
\end{aligned}
\qquad
\begin{aligned}
& h_6 \,  k_1  + h_5  \, k_2  - h_2 \,  k_5 =0  \,,\cr
& h_5 \,  k_1  + h_6  \, k_2  - h_2  \, k_6 =0   \,, 
\end{aligned}
\qquad
\begin{aligned}
& h_6  \, k_4 =0   \,, \cr
& h_6  \, m_4 =0  \,,
\end{aligned}
\label{TriviallyExtendedPoincare_constraints}
\ee
which characterize the complete family of Lie bialgebra structures on the centrally extended (1+1) Poincar\'e algebra~\eqref{bcext}. A glimpse on~\eqref{TriviallyExtendedPoincare_constraints} shows that many solutions exist, but once the coisotropy conditions have been imposed we will be able to single out a smaller number of solutions - those with potential physical interest. 

%


\subsection{Coisotropy with respect to the Lorentz subalgebra $so(1,1)$}

The coisotropy condition~(\ref{so11coisotropy}) with respect to the Lorentz subalgebra generated by $K$ implies that the terms $P_1 \wedge P_0$, $P_1 \wedge \Q$ and $P_1 \wedge \Q$ in $\delta (K)$ in~\eqref{Pre-cocommutator_TriviallyExtendedPoincare} need to vanish, which is tantamount to say 
\be
k_4 =k_5 =k_6 =0 \,.
\label{TriviallyExtendedPoincare_so11coisotropy_constraints}
\ee
As expected, these constraints ensure that $\{x^0,x^1,\q\}$ generate a Lie subalgebra within~\eqref{dual11}, which will be our generic extended noncommutative spacetime, namely
\be
\begin{aligned}
& [x^0,x^1]= k_1 \,x^0 -  k_2 \,x^1  - m_4 \,\q  \,,&
& [x^0,\q]=  h_5  \, x^1+  h_6 \, x^0  \,,&
&  [x^1,\q]=  h_5 \, x^0 +  h_6 \,x^1  \,,&
\end{aligned}
\label{genc}
\ee
which indeed is a non-trivial extension of the algebra~\eqref{genkm11} provided that the constrains~\eqref{TriviallyExtendedPoincare_constraints} are taken into account.

Let us call:
\be
k_1 =a^1 \,, \qquad   k_2 =a^0 \,, \qquad m_4 \, q  = \Theta^{10} = -\Theta^{01}
\ee
so that the first Lie bracket of the dual basis is of the form $[x^\mu , x^\nu] = a^\mu \, x^\nu  - a^\nu \, x^\mu + \Theta^{\mu\nu}$, where $a^\mu$ is a constant vector  and $\Theta^{\mu\nu}$ is an antisymmetric (Lie-algebra-valued) matrix $\Theta^{\mu\nu}$. Moreover, let's call
\be
 h_6  = b_0 \,, \qquad  h_5  =b_1 \,, \qquad  h_2  = \varphi \,,
\ee
Then the (now coisotropic) pre-cocommutator~(\ref{Pre-cocommutator_TriviallyExtendedPoincare}) reads\footnote{Notice that $\epsilon_\mu{}^\nu =\eta_{\mu\rho} \epsilon^{\rho\nu}$, so that $\epsilon_\mu{}^\rho  b_\rho = \delta_\mu{}^0 b_1 + \delta_\mu{}^1b_0$.}
\be
\begin{gathered}
\delta(P_\mu)=  P_\mu\wedge (a^\nu \, P_\nu) + \varphi \,  K \wedge P_\mu
+ b_\mu \, P_0 \wedge \Q +  \epsilon_\mu{}^\nu b_\nu \, P_1 \wedge \Q \,,  \\
\delta(K)= K\wedge (a^\mu \, P_\mu) \,, \qquad  \qquad \delta(\Q)=  m_4 \,P_1\wedge P_0  \,.
\end{gathered}
\ee
where the convention for the Minkowski metric is $\eta_{00} = +1$, $\eta_{11} = -1$ and of course $\eta_{01}=\eta_{10}=0$.
With this notation, the Lie brackets of the extended noncommutative spacetime read
\bea
\begin{aligned}
& [x^0,x^1]= a^1\,x^0 - a^0\,x^1 + \Theta^{01} \,,&  
& [x^0,\q]=  b_0  \, x^0 + b_1  \, x^1  \,,&  
&  [x^1,\q]=  b_1 \, x^0 +  b_0 \,x^1  \,,&
\end{aligned}
\eea
and the rest of the brackets for the dual Lie algebra are
\be
 [\chi ,x^\mu ]= \varphi \, x^\mu + a^\mu\,\chi  \,,\qquad
[\chi,\q]= 0  \,.
\ee
The constraints~(\ref{TriviallyExtendedPoincare_constraints}), after imposing the coisotropy conditions~(\ref{TriviallyExtendedPoincare_so11coisotropy_constraints}), can be written as
\be
b_0 \, a^1 + b_1 \, a^0 = 0 \,, \qquad b_1 \, a^1 + b_0 \ a^0 = 0 \,, \qquad \varphi \, \Theta^{10} =0 \,, \qquad b_0 \, \Theta^{10} = 0 \,.
\label{TriviallyExtendedPoincare_constraints_so11coisotropycase}
\ee
We now study in full generality the solutions to these equations. We have to distinguish two main cases (\textbf{A} and \textbf{B}), which in turn divide into a few sub-cases. The numerical vector $a^\mu$ and the antisymmetric (Lie-algebra-valued) matrix $\Theta^{\mu\nu}$ will allow to interpret the dual algebra as, respectively, $\kappa$-vector-like and canonical $\theta$-noncommutative spacetimes.

\subsubsection*{Case A. Vanishing $\Theta^{10}$ parameter.}

If $m_4 = 0$ and therefore  $\Theta^{10}=0$ we are only left with the first two of Eqs.~(\ref{TriviallyExtendedPoincare_constraints_so11coisotropycase}). There are three sub-cases:
\begin{enumerate}
\item[\textbf{A.1}] If $a^\mu = 0$, then $b_0$ and $b_1$ are two free parameters, thus giving rise to a 3-parameter family of solutions ($b_0,b_1,\varphi$);
\be
\leftBR{
&\delta(K)=0 \,, \cr
&\delta(P_\mu)=\varphi \,  K \wedge P_\mu
+ b_\mu \, P_0 \wedge \Q + \epsilon_\mu{}^\nu b_\nu \, P_1 \wedge \Q  \,, \cr
& \delta(\Q)= 0 \,.
}
\ee
We have a {\em non-central extension by $\q$} of a {\em commutative} Minkowski spacetime:
\be
\begin{aligned}
& [x^0,x^1] = 0 \,,&
& [x^0,\q]=  b_0  \, x^0 + b_1  \, x^1 \,,&  
&  [x^1,\q]=  b_1 \, x^0 +  b_0 \,x^1 \,.&
\end{aligned}
\ee

\item[\textbf{A.2}]  \label{abLightlikeCase}
If $a^\mu$ is lightlike, $a^\mu a_\mu =0$, then  the vector $b_\mu$ is lightlike too and provides only one free parameter. Using `light-cone coordinates' $v^\pm = \frac{1}{\sqrt 2} \left( v^0 \pm v^1 \right)$ the two constraints can be written
\be
b_0 = - \frac{a^0}{a^1} b_1 \,,
\ee
so we see that $a^\pm =0$ implies $b^\mp =0$. The two vectors can be parametrized by their time component, and if we write $a^\mu = a^0 (1, \pm 1)$ that implies  $b_\mu = b_0 (1, \mp 1)$. Therefore we have the following couple  ($a^0,b_0,\varphi$) of 3-parameter families of cocommutators:
\be
\leftBR{
&\delta(K)= a^0 \, K\wedge (P_0 \pm P_1) \,, \cr
&\delta(P_0)=  \varphi \,K\wedge P_0  \mp a^0 \,P_1\wedge P_0 + b_0 (P_0 \mp P_1) \wedge \Q \,, \cr
&\delta(P_1)=  \varphi \,K\wedge P_1  + a^0 \,P_1\wedge P_0 + b_0 (P_1 \mp P_0) \wedge \Q \,, \cr
&\delta(\Q)= 0 \,,
}
\ee
each of which gives rise to a {\em non-central extension by $q$} of a {\em non-commutative} Minkowskian spacetime:
\be
\begin{aligned}
& [x^0,x^1]= a^0 (\pm x^0 + x^1) \,,&
& [x^0,\q]=   b_0 ( x^0 \mp x^1) \,,&  
&  [x^1,\q]=  b_0 ( \mp x^0 + x^1)  \,,&
\end{aligned}
\ee

\item[\textbf{A.3}] 
Finally, if $a^\mu$ is time- or space-like, then $b_\mu = 0$ and we are left with this 3-parameter ($a^0,a^1,\varphi$) family of deformations:
\bea
\leftBR{
&\delta(K)= K\wedge (a^\mu \, P_\mu) \,, \cr
&\delta(P_\mu)=  P_\mu\wedge (a^\nu \, P_\nu) + \varphi \,  K \wedge P_\mu \,, \cr
&\delta(\Q)= 0  \,, 
}
\eea
whose associated noncommutative spacetime is a {\em trivial central extension of the $\kappa$-Minkowski spacetime}:
\be
[x^0,x^1]= a^1\,x^0 - a^0\,x^1\,,   \qquad
[x^\mu,\q]= 0 \,, 
\ee
which would be the usual type of structure arising within a quantum principal bundle construction, where the quantum gauge group and the noncommutative base space commute between themselves.

\end{enumerate}

\subsubsection*{Case B. Nonzero `canonical noncommutativity parameter' $\Theta^{10}$}

If  $m_4 \neq 0$ and therefore $\Theta^{10} \neq 0$, the constraints~(\ref{TriviallyExtendedPoincare_constraints_so11coisotropycase}) imply that $\varphi = b_0 =0$. Then the remaining constraints reduce to
\be
b_1 \, a^0 = 0 \,, \qquad b_1 \, a^1 = 0 \,.
\ee
In this situation we have only the following two cases:
\begin{enumerate}
\item[\textbf{B.1}]  When $a^\mu =0$, we have a 2-parameter family in which  only $\Theta^{10}$ and $b_1$ are nonzero:
\be
\leftBR{
&\delta(K)= 0 \,, \cr
&\delta(P_0) = b_1 \, P_1 \wedge \Q \,, \cr
&\delta(P_1) = b_1 \, P_0 \wedge \Q \,, \cr
&\delta(\Q)=  m_4 \,P_1\wedge P_0  \,, }
\ee
and the associated noncommutative spacetime reads
\be
 [x^0,x^1]=  \Theta^{01}  \,,   \qquad
  [x^0,\q]=  b_1  \, x^1 \,, \qquad
    [x^1,\q]=  b_1 \, x^0 \,.
    \label{theta1}
\ee
When both parameters are different from zero, the Lie algebra~\eqref{theta1}  is isomorphic to the $sl(2)$ Lie algebra. The $b_1= 0$ case leads to the well-known $\theta$-noncommutative spacetime in (1+1) dimensions (see~\cite{DFR, szabo, Aschieri2017} and references therein), which can be thus considered as an extended noncommutative Minkowski spacetime which is invariant under a certain quantum deformation of the extended (1+1) Poincar\'e group. 

\item[\textbf{B.2}] When $b_1=0$,  we obtain a 3-parameter family of  deformations
\bea
\leftBR{
&\delta(K)= K\wedge (a^\mu \, P_\mu) \,, \cr
&\delta(P_\mu)=  P_\mu\wedge (a^\nu \, P_\nu) \,,   \cr
&\delta(\Q)=  m_4 \,P_1\wedge P_0 \,, 
}
\eea
which induces the most general kind of spacetime noncommutativity (vector-like $\kappa$ term plus $\Theta$-term), namely:
\be
 [x^0,x^1]= a^1\,x^0 - a^0\,x^1 + \Theta^{01} \,,     \qquad
 [x^\mu,\q]=   0  \,.
\ee
Note that when $a^\mu\neq 0$ we have a non-trivial central extension of the vector-like $\kappa$-deformation, while in the case $a^\mu =  0$ we recover again the $\theta$-deformation.

\end{enumerate}

%
%
%
%
%
%


\subsection{Coisotropy with respect to $so(1,1) \oplus so(2)$}

We could also consider as the isotropy subgroup for the homogeneous space the central extension of the Lorentz subgroup generated by $I=\mbox{Span}\{K,\Q\}$. In this case the coisotropy condition implies that
\be
\delta(I)\subset I\wedge X, \qquad
I=\mbox{Span}\{K,\Q\},
\qquad X\in {g}.
\ee
and this condition will guarantee that the space and time translation coordinates will close a Lie subalgebra independently of the $\q$ generator (therefore, we will not have an extended noncommutative spacetime).
The corresponding conditions on the generic precocommutator~\eqref{Pre-cocommutator_TriviallyExtendedPoincare} are
\be
k_4 =0, \qquad 
m_4 =0,
\ee
and we obtain
\be
\leftBR{
&\delta(K)= k_1 \,K\wedge P_1+  k_2 \,K\wedge P_0  + k_5 \,P_1\wedge \Q +  k_6 \,P_0\wedge \Q \,, \cr
&\delta(P_0)=  h_2 \,K\wedge P_0  -k_1 \,P_1\wedge P_0 + h_5 \,P_1\wedge \Q +  h_6 \,P_0\wedge \Q \,,\cr
&\delta(P_1)= h_2 \,K\wedge P_1  + k_2 \,P_1\wedge P_0 + h_6 \,P_1\wedge \Q +  h_5 \,P_0\wedge \Q \,,\cr
&\delta(\Q)=  0  \,,
}
\ee
together with two further constrains coming from Jacobi identities:
\be
\begin{aligned}
& h_6 \,  k_1  + h_5  \, k_2  - h_2 \,  k_5 =0 \,,&
& h_5 \,  k_1  + h_6  \, k_2  - h_2  \, k_6 =0 \,.&
\end{aligned}
\ee

Calling $k_6 =c^0$ and  $k_5 =c^1$ we get
\be
\leftBR{
&\delta(K)= K\wedge (a^\mu \, P_\mu)  + (c^\mu P_\mu) \wedge \Q \,, \cr
&\delta(P_\mu)=  P_\mu\wedge (a^\nu \, P_\nu) + \varphi \,  K \wedge P_\mu
+ b_\mu \, P_0 \wedge \Q + \eta_{\mu\rho} \epsilon^{\rho\sigma} b_\sigma \, P_1 \wedge \Q \,,  \cr
&\delta(\Q)=  0  \,,
}
\ee
and the commutators of the dual Lie algebra read
\bea
\begin{aligned}
& [x^0,x^1]= a^1\,x^0 - a^0\,x^1 \,,& & [\chi ,x^\mu ]= \varphi \, x^\mu + a^\mu\,\chi \,,      \cr
& [x^0,\q]=  b_0  \, x^0 + b_1  \, x^1  + c^0 \, \chi \,,& & [\chi,\q]= 0 \,,  \cr  
&  [x^1,\q]=  b_1 \, x^0 +  b_0 \,x^1 + c^1 \, \chi  \,,& &
\end{aligned}
\eea
together with the constraints
\be
b_0\, a^1 + b_1 \, a^0 - \varphi \,  c^1 =0   \,, \qquad b_1\,  a^1 + b_0\, a^0 - \varphi \, c^0=0  \,,
\ee
that by using light-cone coordinates can be written as
\be \textstyle 
b_+ \, a^+ = \frac{1}{\sqrt 2} \varphi \, c^+ \,,
\qquad
b_- \, a^- = \frac{1}{\sqrt 2} \varphi \, c^- \,.
\ee
The solutions of the above equations will be split into the cases when $\varphi$ vanishes and when it doesn't. Note that, as it was expected from the coisotropy condition, $\{x^0,x^1\}$ always generates a Lie subalgebra, and its commutator
\be
[x^0,x^1]= a^1\,x^0 - a^0\,x^1,
\label{coisotr2}
\ee
is just the generalized $\kappa$-Minkowski spacetime.

\section{Extended (1+1) noncommutative (A)dS spacetimes}\label{Sec(A)dS}

It seems natural to wonder which kind of extended noncommutative spacetimes can be obtained when the cosmological constant is non-vanishing, {\em i.e.}, by considering centrally extended (A)dS Lie bialgebras. This analysis can be performed on the same footing as before, by taking into account that the Lie brackets and Casimir operators for the centrally extended (A)dS algebra are given by
\be
[K,P_0]=P_1 \qquad [K,P_1]=P_0 \qquad [P_0,P_1]= - \Lambda \, K \qquad [\Q,\cdot\,]=0 \,,
\label{bc}
\ee
\be
{\cal C}_1= P_0^2 - P_1^2 - \Lambda \, K^2 \,, \qquad {\cal C}_2= M  \,,
\label{bd}
\ee
where the case of $\Lambda>0$ corresponds to the dS case ($\Lambda<0$ is the AdS one), and the limit $\Lambda\to 0$ leads to the centrally extended Poincar\'e algebra~\eqref{bcext}.

After imposing the cocycle condition on the general ansatz~(\ref{Generalcocommutator_TriviallyExtendedPoincare}) we get
\be
\leftBR{
&\delta(K)=  k_1 \,K\wedge P_1+  k_2 \,K\wedge P_0 + k_5 \,P_1\wedge \Q +  k_6 \,P_0\wedge \Q  \,, \cr 
& \delta(P_0) = h_2 \,K\wedge P_0   + h_5 \,P_1\wedge \Q -k_1 \,P_1\wedge P_0  -\Lambda \, k_6 K \wedge \Q  \,, \cr 
&\delta(P_1)=  \Lambda \, k_5 \,K\wedge \Q +  k_2 \,P_1\wedge P_0 + h_2 \,K\wedge P_1 +  h_5 \,P_0\wedge \Q \,, \cr
&\delta(\Q)= 0 \,,
}
\label{adsgen}
\ee
therefore we immediately see that the central generator $\Q$ always has vanishing cocommutator, which immediately precludes the existence of the $\theta$-noncommutativity in any of the extended noncommutative (A)dS spacetimes. Notice also that the solution above is more restrictive than the solution one obtains in the Poincar\'e $\Lambda=0$ case~(\ref{Pre-cocommutator_TriviallyExtendedPoincare}), because some equations following from the cocycle condition vanish altogether when $\Lambda \to 0$. For instance,  the solution above reduces to the solution~(\ref{Pre-cocommutator_TriviallyExtendedPoincare}) in the $\Lambda \to 0$ limit when $m_4 = k_4 = h_6 =0$.

Thus, the dual Lie algebra to the cocommutator~\eqref{adsgen} is
\be
\begin{aligned}
& [x^0,x^1]= k_1 \,x^0 -  k_2 \,x^1   \,,  \cr
& [x^1,\q]=  h_5 \, x^0 + k_5 \,\chi  \,, \cr
& [x^0,\q]=  h_5  \, x^1+  k_6 \,\chi  \,,
\end{aligned}
\qquad
\begin{aligned}
& [x^1,\chi]= -  k_1 \,\chi -h_2 \,x^1  \,,\cr
&  [x^0,\chi]=- k_2 \,\chi + h_2 \,x^0  \,,\cr
&   [\q,\chi]= \Lambda \,  k_6 \, x^0 - \Lambda \,   k_5 \,x^1   \,,
\end{aligned}
\ee
and the Jacobi identities lead to the conditions:
\be
h_5 \, k_2  - h_2 \,  k_5  =0 \,, \qquad  h_5  \, k_1 - h_2  \, k_6   = 0  \,, \qquad  \Lambda \left(  k_2 \,  k_6  -k_1 \,  k_5 \right) = 0 \,,
\ee
which are more restrictive than in the Poincar\'e case, as in the $\Lambda \to 0$ limit the last equation collapses to a tautology. For $\Lambda \neq 0$, the equations above can be rewritten in the following form:
\be
\vec a \times \vec b   =0 \,, \qquad  \vec a \times \vec c  = 0  \,, \qquad   \vec b \times \vec c = 0 \,, \qquad
\left\{ \begin{array}{l}
\vec a = (h_5 , h_2) 
\\
\vec b = (k_5 , k_2) 
\\
\vec c = (k_6 , k_1) 
\end{array}\right. \,,
\label{paralel}
\ee
where the planar Euclidean vector product is defined as $\vec u \times \vec v = u_1 v_2 - u_2 v_1$, and its vanishing implies that the two 2D vectors are parallel.  Therefore the equations~\eqref{paralel} imply that all three vectors are nonzero and parallel to each other:
\be
\vec a = a \left( \cos \theta , \sin \theta \right) \,, \qquad \vec b = b \left( \cos \theta , \sin \theta \right)  \,, \qquad \vec c = c \left( \cos \theta , \sin \theta \right) \, .
\ee
Thus, they can be parametrized by their magnitudes $a$, $b$ and $c$ (which could be zero), and by their common direction angle $\theta$.

In this case too, the extended (1+1) dimensional (A)dS spacetimes $M$ are obtained as homogeneous spaces by taking the Lorentz subgroup generated by $K$ as the invariance subgroup of the origin. Therefore, if we impose the coisotropy condition with respect to $K$ onto the cocommutator~\eqref{adsgen} we get that $k_5 =k_6 =0$, which leads to
\be
h_5 \, k_2  = h_5  \, k_1  =0 \,,
\ee
[which means that $\theta = \frac \pi 2$ in~\eqref{paralel}], and  we have two cases:
\begin{itemize}
\item[\textbf{C.1}]  When $k_1 =k_2 =0$ and $h_5$, $h_2$ are arbitrary we have
\be\label{C1cocomm}
\leftBR{
&\delta(K)=  0 \,, \cr 
& \delta(P_0) = h_2 \,K\wedge P_0   + h_5 \,P_1\wedge \Q  \,, \cr 
&\delta(P_1)=    h_2 \,K\wedge P_1 +  h_5 \,P_0\wedge \Q \,, \cr
&\delta(\Q)= 0 \,,
}
\ee
which leads to the extended noncommutative spacetime
\be
 [x^0,x^1]= 0 \,,  \qquad
 [x^1,\q]=  h_5 \, x^0\,, \qquad
 [x^0,\q]=  h_5  \, x^1 \,.
 \label{ncads11}
\ee
When $h_5\neq 0$ we have a non-central extension of a commutative spacetime, that would correspond in the Poincar\'e case  to the space {\bf A.1} with $b_0=1$.

\item[\textbf{C.2}] When $h_5 = 0$ and  $k_1$, $k_2$, $h_2$ are arbitrary we get
\be\label{C2cocomm}
\leftBR{
&\delta(K)=  k_1 \,K\wedge P_1+  k_2 \,K\wedge P_0  \,, \cr 
& \delta(P_0) = h_2 \,K\wedge P_0   -k_1 \,P_1\wedge P_0   \,, \cr 
&\delta(P_1)=    k_2 \,P_1\wedge P_0 + h_2 \,K\wedge P_1 \,, \cr
&\delta(\Q)= 0 \,,
}
\ee
and we obtain a centrally extended noncommutative (A)dS spacetime
\be
 [x^0,x^1]= k_1 \,x^0 -  k_2 \,x^1   \,,  \qquad
[x^\mu,\q]=  0 \,. 
 \label{ncads13}
\ee
which corresponds to the case {\bf A.3} in the Poincar\'e classification.

\end{itemize}

It is also worth mentionining that if we impose coisotropy with respect to the extended isotropy subgroup $(K,\Q)$, this condition provides {\em no further constraints} with respect to the general solution~\eqref{adsgen}, and the noncommutative spacetime would be again
\be
[x^0,x^1]= k_1 \,x^0 -  k_2 \,x^1\, ,
\ee
which is exactly the same result~\eqref{coisotr2}. Summarizing, the (A)dS case turns out to be much more restrictive than the Poincar\'e case.

At this point it could seem surprising that the noncommutative (A)dS spacetimes~\eqref{ncads11} and~\eqref{ncads13} do not depend on the cosmological constant $\Lambda$ and, therefore, coincide with the corresponding noncommutative Minkowski spacetimes. Indeed, this is true {\em only at first order}, and higher order contributions depending on $\Lambda$ are expected to appear when the full quantum coproduct $\Delta_z$ is constructed and the all-orders noncommutative spacetime is obtained by applying the full Hopf algebra duality or, alternatively, by quantizing the all-orders Poisson-Lie group whose linearization corresponds to the extended noncommutative spacetimes here presented (see~\cite{RossanoPLB,BHMNsigma,ahep} for explicit examples, including the noncommutative $\kappa$-(A)dS spacetime in (2+1) dimensions, which turns out to be a nonlinear deformation of the $\kappa$-Minkowski spacetime).

\section{Canonical noncommutative spacetime as a quantum space}\label{SecCanonical}

In the case of deformations of the trivial central extension of the  Poincar\'e algebra ($\Lambda = 0$), we have case \textbf{B}, in which the `canonical noncommutativity' generator $\Theta^{10}$ is nonzero. Such case admits the interpretation of the bialgebra of symmetries of the (1+1-dimensional) canonical noncommutative Minkowski spacetime:
\begin{equation}\label{Canonical_commutation_relations}
[x^\mu , x^\nu] = \Theta^{\mu\nu} \,,
\end{equation}
where $\Theta^{\mu\nu}$ commutes with $x^\mu$.
In order for the right-hand side of $[x^0,x^1]$ to take that form and to commute with $x^\mu$, all the parameters of case \textbf{B} other than $m_4$ (recall that $\Theta^{01} = m_4 \, \q$)  have to be put to zero (so we are considering a sub-case of \textbf{B.1}, when $b_1 =0$ or a sub-case of \textbf{B.2}, when $a^\mu = 0$). This is what the coalgebra and its dual Lie algebra look like:
\bea
\leftBR{
&\delta(K)= 0 \,, \cr
&\delta(P_0)=\delta(P_1)=  0 \,,   \cr
&\delta(\Q)=  m_4  \,P_1\wedge P_0 \,, 
}
~~
\begin{aligned}
& [x^0,x^1]=  \Theta^{01} \, \,,     \cr
& [x^\mu,\q]=   [\chi,\q]= 0  \,,  \cr
& [\chi ,x^\mu ]= 0\,.  \cr
\end{aligned}
\label{thetadelta}
\eea
The above Lie bialgebra can be interpreted as the infinitesimal version of a quantum (1+1) extended Poincar\'e group, whose quotient with respect to its Lorentz subgroup gives the noncommutative spacetime~(\ref{Canonical_commutation_relations}). We stress that the non-coboundary nature of the cocommutator $\delta$ in~\eqref{thetadelta} can be straightforwardly proven by taking into account that $\Q$ is central and, therefore, any possible classical $r$-matrix would give $\delta(\Q)=0$ when applying~\eqref{cocom}.  

In fact, we know that $\delta$ provides the first order of the deformed coproduct $\Delta_z$, which in this case can be shown to have no higher order terms. This means that the full coproduct of the quantum (1+1) extended Poincar\'e group is
\bea \label{cototaltheta}
\leftBR{
&\Delta_z(K)= 1\otimes K + K\otimes 1 \,, \cr
&\Delta_z(P_0)=  1\otimes P_0 + P_0\otimes 1 \,,   \cr
&\Delta_z(P_1)=  1\otimes P_1 + P_1\otimes 1 \,,
  \\
&\Delta_z(\Q)= 1\otimes \Q + \Q \otimes 1+  z  \,P_1\wedge P_0 \,, 
}
\eea
where $z=m_4$, and~\eqref{cototaltheta} can be easily shown to be an algebra homomorphism with respect to the {\em undeformed} commutation rules~\eqref{bcext}. Therefore, since the coproduct~\eqref{cototaltheta} has only a first-order deformation, the computation of the full Hopf algebra duality cannot give rise to terms other than $[x^0,x^1]=  \Theta^{01}$, and we can conclude that the (1+1) canonical nocommutative spacetime is a true quantum homogeneous space for the extended Poincare quantum group.

This result seems to be quite interesting, because it is well-known that the noncommutative spacetimes of the form~(\ref{Canonical_commutation_relations}) can be shown to be obtained when a twisting element of the Poincar\'e algebra acts on  the algebra of functions of the commutative Minkowski spacetime (see~\cite{Wess,Aschieri,Chaichian,Koch,Lukitwists} and references therein), which acts covariantly on the algebra of functions generated by the relations~(\ref{Canonical_commutation_relations}). However, the latter twisted symmetry does not allow any rigorous interpretation of the canonical noncommutative spacetime as a quantum homogenous space, in contradistinction to the one that we have just presented.

Our approach allows to understand the canonical noncommutative spacetime~(\ref{Canonical_commutation_relations}) as a Lie algebra of noncommutative coordinates, because the antisymmetric matrix  $\Theta^{\mu\nu}$ on the right-hand side of its commutation relations is promoted to a matrix of central generators. In this way we can see the algebra of $x^\mu$ and $\Theta^{01}$ as a subalgebra of the dual Lie algebra to a Lie bialgebra of symmetries. In this sense we can show that in (1+1) dimensions the canonical noncommutative Minkowski spacetime is genuinely a quantum homogeneous space of a certain quantum extended Poincar\'e group, and the central extension of the latter turns out to be essential since it provides the additional noncommutative coordinate that can be identified with the $\Theta^{01}$ generator.

One could hope that the same structure generalizes to higher dimensions, provided that a dual Lie algebra of the form
\begin{equation}\label{Canonical_Lie_algebra}
[x^\mu , x^\nu] = \Theta^{\mu\nu} \,, \qquad [\Theta^{\mu\nu} , x^\mu] = [ \Theta^{\mu\nu} ,\Theta^{\rho\sigma}]=0
\end{equation}
can be found as a subalgebra of the dual to some Lie bialgebra $\delta$. This conjecture can be checked explicitly for Lie bialgebras $\delta$ of the higher dimensional centrally extended Poincar\'e algebra. The generators $\Theta^{\mu\nu}$ of~(\ref{Canonical_Lie_algebra}) will be dual to a number of new generators $T_{\mu\nu}$ which are all {\em assumed} to be central extensions of a Poincar\'e algebra with the suitable dimension:
\begin{equation}
[T_{\mu\nu}, P_\rho] = [T_{\mu\nu},M_{\rho\sigma}]= [T_{\mu\nu},T_{\rho\sigma}]=0 \,.
\end{equation}
Now, the conditions~(\ref{Canonical_Lie_algebra}) translate into the following assumptions on the cocommutators:
\begin{enumerate}
\item  Terms of the form $P_\mu \wedge P_\nu$ can only appear in $\delta(T_{\mu\nu})$.

\item Terms of the form $P_\mu \wedge T_{\rho\sigma}$ and $T_{\mu\nu} \wedge T_{\rho\sigma}$ cannot appear in any Lie bialgebra cocommutator.
\end{enumerate}
Under such conditions we can compute the most general Lie bialgebra in (2+1) and (3+1) dimensions which is generated by $P_\mu$,  $M_{\mu\nu}$ and  $T_{\mu\nu}$. It turns out that these assumptions are strong enough that the cocycle conditions impose that  $T_{\mu\nu}$ is primitive, \emph{i.e.}  $\delta(T_{\mu\nu}) = 0$, both in (2+1) and (3+1) dimensions. But then, since $P_\mu \wedge P_\nu$ appears only in  $\delta(T_{\mu\nu})$, this implies that $[x^\mu ,x^\nu] =0$, and we end up with a trivial solution.

\section{Outlook and discussion}\label{SecConclusions}

In this paper we have shown that quantum group deformations of centrally extended Lorentzian symmetries can be considered as a possible way to circumvent the no-go CMT, since they allow relaxing the assumption of Leibniz action of symmetry generators on tensor product states. The main conclusion of the CMT is that the spacetime symmetry group and the gauge group, when acting on a QFT, can be unified into a combined structure only in a trivial way, \emph{i.e.} as a direct product of Lie groups. However, the additional structures introduced by quantum group symmetries allow for more general ways in which gauge and spacetime symmetries may combine in a hybrid manner when they act on tensor product states.
In this sense one can conceive, in the framework of quantum groups, a more genuine unification of `internal' and `external' symmetries. As part of a conjoined algebraic structure, gauge and spacetime transformations can `mix up' when acting on tensor product states, making it impossible to make a purely-gauge or a purely-spacetime transformation.

Here we have explored this possibility in the case of (1+1) spacetime dimensions, and of an Abelian, 1-dimensional gauge group [\emph{e.g.}~$SO(2)$]. In the commutative case, the CMT implies that the most general group of symmetries of an $SO(2)$ QFT on a Minkowski background is the direct product $ISO(1,1) \times SO(2)$. We relaxed the assumption of having a Lie algebra of symmetries, into having a Lie bialgebra which is coisotropic w.r.t.~the Lorentz subalgebra (and therefore admits a homogeneous quantum Minkowski spacetime as a quotient). The commutation rules for the spacetime, $x$, and gauge, $q$ coordinates that we obtain belong to the following five cases:
\begin{itemize}
\item[\textbf{A.1}]
$[x,x] = 0$, $[x,q] \subseteq x$: commutative spacetime, gauge coordinate acting on $x$ as vector field;
\item[\textbf{A.2}]
$[x,x] \subseteq x$, $[x,q] \subseteq x$: noncommutative spacetime, gauge coordinate acting on $x$ as vector field;
\item[\textbf{A.3}]
$[x,x] \subseteq x$, $[x,q] =0$: trivial quantum principal bundle on noncommutative spacetime;
\item[\textbf{B.1}]
$[x,x] \subseteq q $, $[x,q] \subseteq x$: canonical-like noncommutative spacetime, with right-hand-side non-primitive;
\item[\textbf{B.2}]
$[x,x] \subseteq q + x$, $[x,q] = 0$:  mixture of canonical and $\kappa$-like noncommutative spacetime.
\end{itemize}
If we want to interpret our starting point, the algebra $iso(1,1) \oplus so(2)$, as the algebra of symmetries of a $U(1)$ gauge theory on Minkowski space, then it is natural to identify the coordinates $\{x^\mu,q\}$ as functions on the (trivial) principal fibre bundle $M \times U(1)$ (where $M$ is Minkowski space), which indeed is the geometrical construction underlying a $U(1)$ gauge theory. In the literature, one can find a notion of \emph{quantum principal bundle}~\cite{BMplb93,BMcmp95,Pflaum,Brzezinski, Durdevic, DurdevicJPA, Aschieri2017, Hajac,Sontz}, but this notion seems to be too restrictive: the gauge group is assumed to close a quantum group on its own [which can only be trivial in the case of $SO(2)$], and the coordinate $\q$ on the gauge group is assumed to commute with the noncommutative coordinates of the base space. This situation is realized only in our case \textbf{A.3}, in which the base space is (generalized) $\kappa$-Minkowski, and the fibre is a standard commutative $U(1)$. But our construction allows for more exotic possibilities, since $\q$ can be non-commuting with respect to the coordinates $x$: in case \textbf{A.2} we have a $\kappa$-Minkowski spacetime with a gauge coordinate that acts on the spacetime coordinates, $[x,q] \subseteq x$, and in case   \textbf{A.1} we have the same thing, this time on a \emph{commutative} spacetime. Cases \textbf{B} are stranger yet: the $\q$ coordinate appears on the right-hand-side of the commutation relation of the $x$, in a term that reminds of the \emph{canonical} noncommutative spacetimes (see below). However, in the case  \textbf{B.1} the coordinate $\q$ does not commute with $x$ and therefore the algebra has nothing to do with the canonical spacetimes. The last case,  \textbf{B.2} is the only one which can be interpreted in some sense as a generalization of the canonical spacetimes (in which the commutation relations are a linear combination of those of $\kappa$-Minkowski and the canonical ones). 

After the main analysis described above, we also considered the possibility of requiring coisotropy w.r.t.~the sum of Lorentz and $so(2)$ subalgebras. This ensures that the spacetime coordinates always close a generalized $\kappa$-Minkowski algebra $[x^0,x^1] = a^1 x^0 - a^0 x^1$, while the commutation rules of $x$ with $q$ in principle can contain the Lorentz group coordinate $\chi$. These commutation rules depend on five real parameters which satisfy two quadratic constraints. In each case the commutators $[x,q]$ include a term proportional to $\chi$. The dual statement is that the cocommutator of the Lorentz generator $K$ contains terms of the kind $\Q \wedge P_\mu$, which mix spacetime and gauge generators.

It was then also natural to extend the previous analysis to centrally extended (1+1)-dimensional (A)dS groups with nonvanishing cosmological constant $\Lambda$. We found that this (as is often the case) is more restrictive than Poincar\'e, and only two classes of solutions emerged: 
\begin{itemize}
\item[\textbf{C.1}]
$[x,x] = 0$, $[x,q] \subseteq x$,
\item[\textbf{C.2}]
$[x,x] \subseteq x$, $[x,q] \subseteq 0$,
\end{itemize}
which tend, in the $\Lambda \to 0$ limit, to, respectively, a subcase of \textbf{A.1} and case \textbf{A.3}. In other words, we found that only \textbf{A.1} and  \textbf{A.3} admit a generalization to curved spacetime.

In summary, with the five extended noncommutative Minkowski spacetimes above described, we revealed a new quantum-algebraic possibility to have putative quantum-gravitational effects that circumvent the limitations of the Coleman--Mandula theorem, which might work even on a commutative spacetime (case \textbf{A.1}). The main message is that the full algebra of functions on the principal bundle $M \times U(1)$ can be noncommutative by assuming  quantum group Poincar\'e symmetries to hold, and can be so in such a way that the simultaneous determination of the gauge coordinate $\q$ and spacetime coordinates $x$ is subject to quantum limitations (because $[x,\q] \neq 0$).  This is highly suggestive of situations in which we have to renounce either to the perfect knowledge of the value of some charged field or of the spacetime coordinate at which it is calculated. However such an interpretation would require a suitable formulation of noncommutative gauge theory on fibre bundles with ``hybrid'' symmetries. Indeed, in order to talk about QFT on a noncommutative spacetime arising from a quantum group symmetry, one needs to accept a form of ``Hopf-algebrization'' of the standard formulation of field theory, in which scalar fields are considered as the element of an algebra which, in the commutative case, is the (Abelian) algebra of functions on the spacetime manifols, and, in the noncommutative case, turns into a non-Abelian one.

Finally, our last exploration involved considering the possibility that the additional coordinate $\q$ has nothing to do with a coordinate on (a neighbourhood of the identity of) the gauge group, and instead it could be interpreted in the context of canonical noncommutative spacetimes. These are well-studied noncommutative spacetimes~\cite{Wess,Aschieri,Chaichian,Koch,Lukitwists} in which the commutator between two spacetime coordinates is equal to a constant times the identity operator: $[x^\mu , x^\nu] = \theta^{\mu\nu} 1$ (much like the Heisenberg commutation relations $[p,q]=i\hbar$). Such spacetimes have interesting properties, \emph{e.g.} the coordinates respect a Heisenberg uncertainty principle~\cite{MoyalArea}, the corresponding QFTs present the phenomenon of IR/UV mixing, and the only known example of QFT defined at all scales, the Grosse--Wulkenhaar model~\cite{GrosseWulkenhaar}, is defined on such a noncommutative background. However, the commutation relations $[x^\mu , x^\nu] = \theta^{\mu\nu} 1$ are not usually considered in a Lie algebra framework, thus precluding the interpretation of $\theta$-spacetimes as quantum homogeneous spaces under a given quantum-group. 

In this paper, we presented an alternative possibility that allows to interpret the (1+1)-dimensional canonical spacetime as the homogeneous space of a quantum Poincar\'e group. This can be achieved by considering the right-hand side of $[x^\mu , x^\nu] = \theta^{\mu\nu}$ as the dual noncommutative coordinate associated to central generator of the extended (1+1) Poincar\'e Lie algebra. In this way, the canonical commutation rule can be obtained as one of the brackets of a dual extended Poincar\'e Lie bialgebra. We then constructed the quantum group associated to this Lie bialgebra, and the coproducts are all primitive except for that of the central generator $\Q$. This approach to describing the symmetries of the canonical spacetime is an alternative one to that of twisting the Poincar\'e group presented in~\cite{Chaichian,Koch,Lukitwists}, and, unlike it, allows to understand the noncommutative spacetime as a quantum homogeneous space. Finally, we tested whether a similar approach could be used in more than (1+1) dimensions. Unfortunately we encountered an obstruction: for $d>1$, there is no bialgebra deformation of the $(d+1)$-dimensional Poicar\'e algebra extended with $d(d+1)/2$ central generators, in which said generators have a nonzero cocommutator. This means that the dual Lie algebra can never be of the form $[x^\mu , x^\nu] = \Theta^{\mu\nu}$ with  $[x^\rho ,  \Theta^{\mu\nu}] =  [\Theta^{\mu\nu},\Theta^{\rho\sigma}]=0$. This shows that only in (1+1) dimensions the canonical noncommutative spacetime can be obtained as a quantum space for a centrally extended Poincar\'e algebra. Nevertheless this does not exclude the possibility of recovering $\theta$-noncommutative spacetimes as duals of Lie bialgebra structures associated to other Lie algebras.

Also, we would like to stress that the approach here presented for the construction of hybrid gauge symmetries can be generalized to both higher dimensional kinematical Lie algebras $t$ of spacetime symmetries and higher dimensional (non-abelian) gauge symmetries with gauge Lie algebra $g$. In such cases, the Lie bialgebra structures $(b,\delta)$ of the direct sum Lie algebra $b=t\oplus g$ will in general intertwine both symmetries, and for those cocommutators $\delta$ that fulfil the coisotropy condition, the corresponding dual algebra $\delta^\ast$ will contain as a subalgebra  the (first order) noncommutative coordinates in the associated noncommutative bundle. Work along these lines is in progress.

\section*{Acknowledgements}

A.B. has been partially supported by Ministerio de Econom\'{i}a, Industria y Competitividad (MINECO, Spain) under grant  MTM2016-79639-P (AEI/FEDER, UE), by Junta de Castilla y Le\'on (Spain) under grant VA057U16 and by the Action MP1405 QSPACE from the European Cooperation in Science and Technology (COST). F.M. was funded by the Euroean Union and the Istituto Italiano di Alta Matematica under a  Marie Curie COFUND action.




\begin{thebibliography}{99}


 
  
\bibitem{ColemanMandula}
S.~Coleman and J.~Mandula,
{{\em Phys. Rev.} {\bfseries 159}, 1251 (1967)}  

\bibitem{Fewster} C. J. Fewster,
{{\em Commun. Math. Phys.} {\bf 357}, 353--378 (2018)}

\bibitem{Carlip2+1book}
S.~Carlip,
{{\em Quantum gravity in 2+1 dimensions}},
Cambridge University Press, (2003)

\bibitem{AT} A. Achucarro  and P. K. Townsend
 {\em Phys.~Lett.} {\bf B180},  89 (1986)

\bibitem{Witten1} E. Witten  {\em Nucl.~Phys.}  {\bf B311}, 46 (1988)

\bibitem{FreidelLivine}
L. Freidel and E. R. Livine
{\em Phys. Rev. Lett.} {\bf 96}, 221301 (2006)

\bibitem{Matschull}  H. J. Matschull, M. Welling, 
{\em Class. Quant. Grav.} {\bf 15} 2981--3030 (1998)

\bibitem{drinfeld87}  V. G. Drinfel'd ,
{\em  Proc. Int. Congress of Math.} 1, 798 (1987),
ed   A. V. Gleason, Berkeley, AMS

\bibitem{CP}  V. Chari and A.  Pressley, {\em A Guide to Quantum Groups}, Cambridge University Press (1994)

\bibitem{MajidBook}
S.~Majid, 
{{\em Foundations of quantum group theory}},
Cambridge University Press (2000)

\bibitem{DrinfeldPL}
V.~Drinfel'd, 
{\em Soviet Math. Dokl.} {\bfseries 27}, 68--71 (1983)

\bibitem{AMII}  A. Y. Alekseev and A. Z. Malkin,
 {\em Commun.~Math.~Phys.} {\bf 169},  99  (1995)

\bibitem{FR} 
V. V. Fock and A. A. Rosly.
{\em Am. Math. Soc. Transl.}  {\bf 191}, 67 (1999)

\bibitem{MSquat} C. Meusburger and B. Schroers,
{\em J. Math. Phys.} {\bf 49}, 083510 (2008)

\bibitem{MS} C. Meusburger and B. Schroers,
{\em Nucl. Phys. B} {\bf 806},  462 (2009)

\bibitem{MatteoFlavioBialgebras} F. Mercati and M. Sergola,
\href{https://arxiv.org/abs/1802.09483}{\em arXiv:1802.09483} (2018)

\bibitem{Lukierskia}
J.  Lukierski, H. Ruegg,   A. Nowicki and V. N.  Tolstoy, 
{\em Phys. Lett.} {\bf B264}, 331   (1991)

 \bibitem{Lukierskib}
 J.  Lukierski,  A. Nowicky and H.  Ruegg,
 {\em Phys. Lett.}  {\bf B271}, 321  (1991)

 \bibitem{kMinkowski}  
P. Maslanka,
{\em J. Phys.} {\bf A26}, L1251 (1993);  \\
S. Majid and H. Ruegg,
{\em Phys. Lett.} {\bf B334}, 348 (1994);\\
S. Zakrzewski,
{\em J. Phys.} {\bf A27}, 2075 (1994); \\
J. Lukierski and H. Ruegg,
{\em Phys. Lett.} {\bf B 329}, 189 (1994)

 \bibitem{BHOS4D} 
A. Ballesteros, F. J. Herranz, M. A. del Olmo and M. Santander,
{\em J. Math. Phys} {\bf 35}, 4928 (1994)

 \bibitem{tmatrix} 
A. Ballesteros, F. J. Herranz, M. A. del Olmo, C. M. Pere\~na and M. Santander,
{\em J. Phys.} {\bf A28}, 7113--7125 (1995)

 \bibitem{nullplane} 
A. Ballesteros, F. J. Herranz, M. A. del Olmo and M. Santander,
{\em Phys. Lett.} {\bf B351}, 137 (1995)

\bibitem{Zakrzewski1997}
S.~Zakrzewski,
{{\em Commun. Math. Phys.} {\bfseries 185}, 285--311 (1997)}
  
  
  
\bibitem{Lukitwists} J. Lukierski and M. Woronowicz,
{\em Phys. Lett.} {\bf B633} 116--124 (2006)

\bibitem{Mandula}
J. E. Mandula, {\em Scholarpedia} {\bf 10}(2):7476, (2015)

\bibitem{HLS}  R. Haag, J. T. Lopuszanski, and M. Sohnius,
{\em Nucl. Phys.} {\bf B88}, 257 (1975)

\bibitem{BernardLeclair}
D. Bernard and A. Le Clair,
{\em Commun. Math. Phys.} {\bf 142}, 99--138 (1991)

\bibitem{Vassilevich} 
D. V. Vassilevich,
{\em Mod. Phys. Lett.} {\bf A21}, 1279--1284 (2006)

\bibitem{CT} M. Chaichian and A. Tureanu,
{\em Phys. Lett.} {\bf B637}, 199 (2006) 

\bibitem{CTZ} M. Chaichian, A. Tureanu and G. Zet,
{\em Phys. Lett.} {\bf B651} 319--323 (2007)


\bibitem{Aschieri} P. Aschieri, M. Dimitrijevic, F. Meyer, S. Schraml and J. Wess,
{\em Lett. Math. Phys.} {\bf 78}, 61--71 (2006)

\bibitem{LO} M. Lindner and S. Ohmer,
{\em Phys. Lett.} {\bf B773}, 231--235 (2017)

 \bibitem{BHOS2D} 
A. Ballesteros, F.J. Herranz, M.A. del Olmo and M. Santander,
{\em J. Phys.} {\bf A26}, 5801--5823 (1993)

\bibitem{azcarrag}
J. A. de Azc\'arraga and J.C. P\'erez Bueno,
{\em J. Math. Phys.} {\bf 36}, 6879--6896 (1995)

\bibitem{Koor} M. S. Dijkhuizen and T. H. Koornwinder,
{\em Geom. Dedicata} {\bf 52}, 291 (1994)

  
\bibitem{Zakrzewski} S. Zakrzewszki, Poisson homogeneous spaces, in: J.
Lukierski, Z. Popowicz, J. Sobczyk (eds.), Quantum Groups (Karpacz 1994), PWN Warszaw (1995) 629


 \bibitem{BMN}
A. Ballesteros, C. Meusburger and  P. Naranjo,
{\em J. Phys.} {\bf A50}, 395202 (2017)

\bibitem{DFR}  S. Doplicher, K. Fredenhagen and J. E. Roberts,
{\em Phys. Lett.} {B331}, 39  (1994); \\
S. Doplicher, K. Fredenhagen and J.E. Roberts,
{\em Commun. Math. Phys.} {\bf 172}, 187 (1995)

\bibitem{szabo} R. J. Szabo,
{\em Phys. Rept.} {\bf 378}, 207-299 (2003)

\bibitem{Aschieri2017} P. Aschieri, P. Bieliavsky, C. Pagani and A. Schenkel, 
{\em Commun. Math. Phys.} {\bf 352}, 287 (2017)

\bibitem{EnricoGalilei} A. Ballesteros, E. Celeghini and F. J. Herranz, 
{\em J. Phys.} {\bf A33}, 3431 (2000)

\bibitem{Opanowicz21} A. Opanowicz,
{\em J. Phys.} {\bf A31}, 2887 (1998)

\bibitem{heis} A. Ballesteros, F.J. Herranz and P. Parashar,
{\em J. Phys.} {\bf A30}, L149 (1997)

\bibitem{NappiWitten} C. Nappi and E. Witten,
{\em Phys. Rev. Lett.} {\bf 71}, 3751 (1993)

\bibitem{CGannals} D. Cangemi and R. Jackiw,
{\em Annals Phys.} {\bf 225}, 229--263 (1993)

\bibitem{RossanoPLB}  
A. Ballesteros, F. J. Herranz and N. R. Bruno,
{\em Phys. Lett.} {\bf B574}, 276 (2003)

\bibitem{BHMNsigma}
A. Ballesteros, F. J. Herranz, C. Meusburger and P. Naranjo,
{\em SIGMA} {\bf 10}, 052 (2014) 

\bibitem{ahep}
A. Ballesteros, N. R. Bruno, F. J. Herranz,
{\em Adv. High Energy Phys.} 7876942 (2017)

\bibitem{Wess} J. Wess,
\href{https://arxiv.org/abs/hep-th/0408080}{\em arXiv:hep-th/0408080} (2004)

\bibitem{Chaichian}  M. Chaichian, P. Kulish, K. Nishijima and A. Tureanu, 
{\em Phys. Lett.} {\bf B604}, 98--102 (2004)

\bibitem{Koch} F. Koch and E. Tsouchnika,
{\em Nucl. Phys.} {\bf B717}, 387--403 (2005)

\bibitem{BMplb93} T. Brzezinski and S. Majid,
{\em Phys. Lett.} {\bf B298}, 339 (1993)

\bibitem{BMcmp95} T. Brzezinski and S. Majid,
{\em Commun. Math. Phys.} {\bf 157},  591 (1993); {\em ibid.} {\bf 167}, 235 (1995) (erratum)

\bibitem{Brzezinski} T. Brzezinski,
{\em J. Geom. Phys.} {\bf 20}, 349 (1996)

\bibitem{Pflaum} M. Pflaum,
{\em Commun. Math. Phys.} {\bf 166}, 279 (1994)

\bibitem{Durdevic} M. Durdevic,
\href{https://arxiv.org/abs/hep-th/9311029}{\em arXiv:hep-th/9311029} (1993)

\bibitem{DurdevicJPA} M. Durdevic,
{\em J. Phys.} {\bf A30}, 2027--2054 (1997)

\bibitem{Hajac} P. M.Hajac,
{\em Commun. Math. Phys.} {\bf 182}, 579 (1996)

\bibitem{Sontz} S. B. Sontz,
{\em Principal Bundles: The Quantum Case}, Springer (2015)

\bibitem{MoyalArea}
G. Amelino-Camelia, G. Gubitosi and F. Mercati,
{\em Phys. Lett.} {\bf B676}, 180--183 (2009)


\bibitem{GrosseWulkenhaar}
H. Grosse and R. Wulkenhaar,
{\em JHEP} {\bf 0312}, 019 (2003);\\
H. Grosse and R. Wulkenhaar,
{\em Commun. Math. Phys.} {\bf 254}, 91--127 (2005);\\
H. Grosse and R. Wulkenhaar,
{\em Lett. Math. Phys.} {\bf 71}, 13 (2005)\\
M, Buric and M. Wohlgenannt,
{\em JHEP} {\bf 1003}, 053 (2010)



\end{thebibliography}
\end{document}